\documentclass[12pt,a4paper]{article} %UKenglish,
\usepackage[utf8]{inputenc}
\usepackage{bm}
\usepackage{color}
\usepackage{amsmath}
\usepackage{amsthm}
\usepackage{amssymb}
\usepackage{graphicx}
\usepackage{verbatim}
\usepackage[breaklinks]{hyperref}
\usepackage{multirow}

\newtheorem{lemma}{Lemma}
\newtheorem{theorem}{Theorem}
\newtheorem{corollary}{Corollary}
\newtheorem{claim}{Claim}

\theoremstyle{definition}
\newtheorem{definition}{Definition}

\newcommand{\set}[2]{\ensuremath{ \{ \, #1 \mid #2 \, \} }}

\renewcommand{\emptyset}{\varnothing}
\renewcommand{\epsilon}{\varepsilon}

\newcommand{\lcm}{\mathop{\mathrm{lcm}}}
\newcommand{\sizere}[1]{|\mathrm{alph}(#1)|}

\begin{document}
\sloppy
\emergencystretch 3em

\title{From regular expressions to deterministic finite automata: $2^{\frac{n}{2}+\sqrt{n}(\log n)^{\Theta(1)}}$ states are necessary and sufficient}
\author{Olga Martynova \and Alexander Okhotin}
\maketitle

\begin{abstract}
It is proved that every regular expression of alphabetic width $n$,
that is, with $n$ occurrences of symbols of the alphabet,
can be transformed into a deterministic finite automaton (DFA)
with $2^{\frac{n}{2}+(\frac{\log_2 e}{2\sqrt{2}}+o(1))\sqrt{n\ln n}}$ states
recognizing the same language
(the best upper bound up to date is $2^n$).
At the same time, it is also shown
that this bound is close to optimal,
namely, that there exist regular expressions of alphabetic width $n$
over a two-symbol alphabet,
such that every DFA for the same language
has at least $2^{\frac{n}{2}+(\sqrt{2} + o(1))\sqrt{\frac{n}{\ln n}}}$ states
(the previously known lower bound is $\frac{5}{4}2^{\frac{n}{2}}$).
The same bounds are obtained for an intermediate problem
of determinizing nondetermistic finite automata (NFA)
with each state having all incoming transitions by the same symbol.
\end{abstract}

\section{Introduction}

Transforming a regular expression of a given size
to DFA with as few states as possible
is a theoretically and practically important problem.
This problem depends on the definition of size of a regular expression.
Several possible measures of size were considered in the literature:
the total number of characters in the string representation of a regular expression;
the number of characters excluding brackets (the reverse Polish length),
and the number of occurences of symbols of the alphabet, not counting brackets or operators
(the alphabetic width).
Ellul, Krawetz, Shallit and Wang~\cite{EllulKrawetzShallitWang} called the alphabetic width ``the most useful in practice''.

Using the alphabetic width as the measure of size,
the problem is formally stated as follows:
how many states are necessary in a DFA
to recognize a language defined by a regular expression of alphabetic width $n$?
In their survey of regular expressions, Ellul et al.~\cite{EllulKrawetzShallitWang}
present this problem as Open Problem 1.
In this paper, the alphabetic width of a regular expression
will be called simply its \emph{size}.

The only known upper bound on the number of states in a DFA
sufficient for representing every regular expression of size $n$
is $2^n+1$ states.
This bound was first proved by Glushkov~\cite[Thm.~16]{Glushkov},
and later it was independently obtained by Leiss~\cite{Leiss,Leiss2},
who has at the same time established a lower bound of the order $\Omega(2^{\frac{n}{3}})$.
Leung~\cite[\S{}3]{Leung1998} proved a lower bound of $2^{\frac{n}{2}}$ states,
which was improved to $\frac{5}{4} \cdot 2^{\frac{n}{2}}$
by Ellul et al.~\cite[Thm.~11]{EllulKrawetzShallitWang}.
These state-of-the-art bounds are mentioned
in a survey by Gruber and Holzer~\cite[Thm.~11]{GruberHolzer_survey}.

Clearly, the known lower bound $\frac{5}{4} \cdot 2^{\frac{n}{2}}$
and upper bound $2^n+1$ are quite far apart!
In this paper, a substantially more efficient transformation
from a regular expression of size $n$
to a DFA will be proved,
which will use only $2^{\frac{n}{2}+(\frac{\log_2 e}{2\sqrt{2}}+o(1))\sqrt{n\ln n}}$ states.
At the same time, the lower bound will also be improved
to $2^{\frac{n}{2}+(\sqrt{2} + o(1))\sqrt{\frac{n}{\ln n}}}$ states.
Thus it will be proved
that the necessary and sufficient number of states
is of the order $2^{\frac{n}{2}+\sqrt{n}(\log n)^{\Theta(1)}}$.

Besides the bounds on the size of DFA described above,
there are a lot of other results on different transformations of regular expressions to automata.
The most well-known is the classical Thompson's~\cite{Thompson} construction,
which converts a regular expression to an NFA with $\epsilon$-transitions ($\epsilon$-NFA);
it is a convenient first step in a transformation to a DFA.
Ilie and Yu~\cite{IlieYu} defined an alternative transformation
of a regular expression to an $\epsilon$-NFA
that uses fewer transitions than Thompson's;
assuming that the size of an automaton is the sum of the number of states and the number of transitions,
they obtained a lower bound of $4n-1$ and an upper bound of $9n-\frac{1}{2}$.
Later, Gruber and Gulan~\cite{GruberGulan} established a precise bound of $4.4n+1$.

The complexity of direct transformation of regular expressions to NFA,
bypassing the $\epsilon$-NFA stage and aiming to minimize the number of transitions,
has also been studied.
Hromkovi\v{c} et al.~\cite{HromkovicSeibertWilke}
reduced the number of transitions from the obvious $O(n^2)$ to $O(n (\log n)^2)$,
and also presented an example of a regular expression,
for which every automaton needs to have at least $\Omega(n \log n)$ transitions.
Hagenah and Muscholl~\cite{HagenahMuscholl} developed a faster algorithm
for doing this transformation.
Lifshits~\cite{Lifshits} improved the lower bound on the number of transitions
to $\Omega\big(\frac{n (\log n)^2}{\log\log n}\big)$.
Finally, Schnitger~\cite{Schnitger} proved the lower bound of $\Omega(n (\log n)^2)$ transitions,
which asymptotically matches the upper bound by Hromkovi\v{c} et al.~\cite{HromkovicSeibertWilke}.

The case of a unary alphabet was investigated separately.
Ellul et al.~\cite[Thm.~14]{EllulKrawetzShallitWang} proved
that every unary regular expression of size $n$
can be transformed to a DFA with at most $g(n)+(n-1)^2+2$ states,
where $g(n)$ is Landau's function, which is asymptotically of the order $e^{(1+o(1))\sqrt{n \ln n}}$.
Also they proved a lower bound of $g(n)$ states.

Something is known on the complexity of transforming automata back to regular expressions.
For the classical transformation of McNaughton and Yamada~\cite{McnaughtonYamada}
Ellul et al.~\cite[Thm.~17]{EllulKrawetzShallitWang} proved
that an NFA with $n$ states is transformed to a regular expression
of size at most $|\Sigma| \cdot n 4^n$.
Ehrenfeucht and Zeigler~\cite{EhrenfeuchtZeigler} showed that a DFA with $n$ states
may require a regular expression of size at least $2^{n-1}$;
this lower bound was proved for an alphabet that grows quadratically in $n$.
Gruber and Holzer~\cite{GruberHolzer_dfa_to_reg_expr_lower_bound},
obtained a lower bound of $2^{\Theta(n)}$ using a binary alphabet.
Also Gruber and Holzer~\cite{GruberHolzer_dfa_to_reg_expr_upper_bound}
improved the upper bound in the case of a fixed alphabet;
in particular, by their method, every DFA with $n$ states over a two-symbol alphabet
is transformed to a regular expression of size $O(1.742^n)$.
In the case of a unary alphabet, Martinez~\cite{Martinez} proved an upper bound of $O(n^2)$,
whereas Gawrychowski~\cite{Gawrychowski} improved it to $O(\frac{n^2}{(\log n)^2})$.

In this paper, the proposed improved upper bound
on the transformation from a regular expression to a DFA
is proved in two stages.
At the first stage, described in Section~\ref{section_nfa_remembering_last_symbol},
a regular expression is transformed to an intermediate model:
an \emph{NFA that remembers the last symbol it has read}.
Next, this NFA is transformed to a DFA by the subset construction,
and in Section~\ref{section_upper_bound},
the number of reachable subsets is estimated.
This number gives the upper bound.
Finally, in Section~\ref{section_lower_bound},
an example of a regular expression is constructed,
so that its transformation to a DFA requires many states;
the same language is also defined by an NFA that remembers the last symbol,
so that the lower bound also applies
to the transformation of NFA of this kind to DFA.

\section{Regular expressions and finite automata}

We consider standard regular expressions.
\begin{definition}
Let $\Sigma$ be a finite alphabet.
Regular expressions over the alphabet $\Sigma$
and the languages they describe
are defined inductively as follows.
\begin{itemize}
\item
	A symbol $a \in \Sigma$ is a regular expression,
	which defines the language $L(a) = \{a\}$.
\item
	The empty set $\emptyset$ is a regular expression,
	with $L(\emptyset) = \emptyset$.
\item
	A concatenation $(\alpha\beta)$ of two regular expressions $\alpha$ and $\beta$
	is a regular expression
	that defines
	$L((\alpha\beta)) = \set{uv}{u\in L(\alpha), v \in L(\beta)}$.
\item
	A disjunction $(\alpha\ | \ \beta)$ of two regular expressions $\alpha$ and $\beta$
	is a regular expression defining the union of their languages:
	$L((\alpha\ | \ \beta)) = L(\alpha) \cup L(\beta)$.
\item
	A Kleene star $(\alpha)^*$ of a regular expression $\alpha$ is also a regular expression,
	with $L((\alpha)^*) = \set{u_1u_2\ldots u_k}{k \in \mathbb{N} \cup \{0\},\; 
	u_1,\ldots,u_k \in L(\alpha)}$.
\end{itemize}
\end{definition}

Brackets can be omitted, with the following default precedence:
first Kleene star, then concatenation, then disjunction.

\begin{definition}
The alphabetic width of a regular expression $\alpha$,
denoted by $\sizere{\alpha}$,
is the number of occurrences of all symbols of the alphabet in $\alpha$,
counted with multiplicity.
\end{definition}

In this paper, alphabetic width is called simply \emph{size},
as no other succinctness measures are discussed.

Finite automata, deterministic and nondeterministic,
are also defined in the standard way.
\begin{definition}
A nondeterministic finite automaton (NFA) is a quintuple $A=(\Sigma, Q, Q_0, \delta, F)$,
where
\begin{itemize}
\item
	$\Sigma$ is a finite input alphabet,
\item
	$Q$ is a finite set of states,
\item
	$Q_0 \subseteq Q$ is the set of initial states,
\item
	$\delta \colon Q \times \Sigma \to 2^Q$ is the transition function,
\item
	$F \subseteq Q$ is the set of accepting states.
\end{itemize}
A computation of $A$ on a string $w=a_1 \ldots a_m$
is a sequence of states $p_0, p_1, \ldots, p_m$,
with $p_0 \in Q_0$ and $p_{i+1} \in \delta(p_i, a_{i+1})$ for all $i$.
A computation is accepting if $p_m \in F$.
The language defined by the automaton
is the set of all strings with at least one accepting computation,
and it is denoted by $L(A)$.

A deterministic finite automaton (DFA)
is an NFA
with $|Q_0|=1$ and $|\delta(q, a)|=1$ for all $q \in Q$ and $a \in \Sigma$.
\end{definition}

Every NFA $A=(\Sigma, Q, Q_0, \delta, F)$
can be transformed to a DFA by the \emph{subset construction}:
the resulting DFA has the set of states $2^Q$,
with initial state $Q_0 \in 2^Q$,
transition function $\delta'(S, a)=\bigcup_{q \in S} \delta(q, a)$,
and accepting states $F'=\set{S \subseteq Q}{S \cap F \neq \emptyset}$.

\section{An intermediate model}\label{section_nfa_remembering_last_symbol}

The classical transformation from regular expressions to DFA
first transforms them to NFA, and then to DFA by the subset construction.
It turns out that if the first step (the transformation to NFA)
is made carefully,
then one can obtain an NFA of the following special form.

\begin{definition}
An NFA $A=(\Sigma, Q, q_0, \delta, F)$
is said to \emph{remember the last symbol},
if its set of states splits into disjoint subsets $Q_a$, for all $a \in \Sigma$,
so that every state from $Q_a$ can be entered only by the symbol $a$,
that is, $\delta(q, a) \subseteq Q_a$ for all $q \in Q$ and $a \in \Sigma$.
\end{definition}

\begin{lemma}\label{regular_expression_to_nfa_lemma}
For every regular expression of size $n$,
there is an NFA with $n+1$ states that remembers the last symbol
and recognizes the same language.
\end{lemma}

Actually, the NFA constructed by Leiss~\cite{Leiss,Leiss2}
does remember the last symbol,
even though this property is not considered in the cited papers.

\begin{proof}
The automaton constructed from a regular expression
will have an extra property:
it will have a unique initial state with no incoming transitions.

The proof is by induction on the structure of a regular expression;
all five cases are illustrated in Figure~\ref{f:regular_expression_to_nfa}.

\begin{figure}[t]
	\centerline{\includegraphics{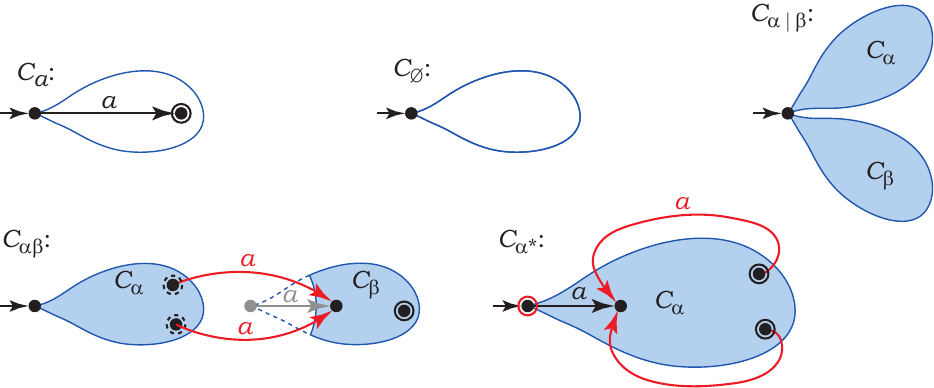}}
	\caption{Transformation of a regular expression to an NFA that remembers the last symbol
		and has a non-reenterable initial state.}
	\label{f:regular_expression_to_nfa}
\end{figure}

Base case I: regular expression $a$, for $a \in \Sigma$.
An automaton with $2$ states and one transition by $a$ is constructed.

Base case II: regular expression $\emptyset$.
The automaton has $1$ state and no transitions.

Induction step, case I: regular expression $\alpha \ | \ \beta$.
Let $C_\alpha$ and $C_\beta$ be the automata
constructed for $\alpha$ and $\beta$ by the induction hypothesis.
Then the automaton $C_{\alpha \ | \ \beta}$ is obtained as a union of these two automata,
preserving all transitions and accepting states, and with initial states joined into one.
The joint initial state will be accepting if and only if
at least one of the initial states of $C_\alpha$ and $C_\beta$ was accepting.
Since no new transitions have been added,
and there are no transitions to the joint initial state,
the new automaton remembers the last symbol as well.
Since $C_\alpha$ has $\sizere{\alpha}+1$ states
and $C_\beta$ has $\sizere{\beta}+1$ states,
the constructed automaton will have
$\sizere{\alpha}+1+\sizere{\beta}+1-1=\sizere{\alpha \ | \ \beta} + 1$ states.

Induction step, case II: regular expression $\alpha \beta$.
Let $C_\alpha$ and $C_\beta$ be the automata constructed for $\alpha$ and $\beta$.
Then the new automaton is obtained as follows.
Its set of states is the disjoint union of the sets of states of $C_\alpha$ and $C_\beta$,
with the initial state of $C_\beta$ deleted.
For each transition from the initial state of $C_\beta$ by any symbol $a$ to any state $q$,
the new automaton has transitions by the same symbol and to the same state 
for every accepting state of $C_\alpha$.
The accepting states of the new automaton are all the accepting states $C_\beta$,
and also, if $C_\beta$ accepts the empty string,
then all accepting states of $C_\alpha$ are made accepting.
The resulting automaton remembers the last symbol,
because all transitions added lead to their destination states
by the same symbols as the deleted transitions to these states.
The number of states in the automaton constructed is
$\sizere{\alpha}+1+\sizere{\beta}+1-1=\sizere{\alpha \ | \ \beta} + 1$.

Induction step, case III: regular expression $\alpha^*$.
Let $C_\alpha$ be the NFA for $\alpha$.
The new automaton is obtained from $C_\alpha$ by reconstructing it in the following way.
For each transition from the initial state to some state $q$ by some symbol $a$,
transitions from each accepting state to the state $q$ by $a$ are added.
All accepting states of $C_\alpha$ remain accepting,
and the initial state also becomes accepting.
The automaton remembers the last symbol,
because all new transitions lead to their destination states
by the same symbols as the existing transitions.
The number of states and the size of the regular expression remain unchanged.
\end{proof}

\section{Upper bound}\label{section_upper_bound}

In order to prove an upper bound on the size of a DFA
recognizing the language of a regular expression of a given size,
such a bound will first be proved for determinization
of NFA that remember the last symbol.
Then, due to Lemma~\ref{regular_expression_to_nfa_lemma},
the bound for the case of regular expressions will follow immediately.

The notion of an NFA that remembers the last symbol is beneficial,
because when the subset construction is applied to such an automaton,
a lot of subsets turn out to be unreachable.
Except for the initial subset,
any other reachable subset consists of states
that are enterable by the same symbol.
It will be proved that the overall number of subsets
is substantially smaller
than the number of states necessary for determinizing NFA of the general form.

The plan for proving an upper bound
is to calculate the number of reachable subsets by two different methods,
and to take the minimal value in each case.
The first calculation method
gives a good upper bound
in the case when the original NFA has no ``dominating symbol'',
by which more than half of the states would be reachable
(actually, a little bit more than a half).

\begin{lemma}[The first method for subset calculation]
Let $A=(\Sigma, Q, Q_0, \delta, F)$
be an $n$-state NFA that remembers the last symbol.
For each symbol $a \in \Sigma$,
let $Q_a \subseteq Q$ be the subset of states reachable by $a$.
Let $n_1$ be the maximal number of states reachable by the same symbol:
$n_1 = \max_{a \in \Sigma} |Q_a|$.
Then there exists a DFA with at most
$\max(2^{\frac{n}{2}+1}, 2^{n_1+1})$
states that recognizes the same language.
\end{lemma}
\begin{proof}
The standard subset construction is used,
and the number of reachable subsets is estimated from above.
Let $\Sigma = \{a_1,\ldots, a_k\}$,
and let the symbols be ordered by the number of states reachable by these symbols:
$|Q_{a_1}| \geqslant \ldots \geqslant |Q_{a_k}| \geqslant 0$.
Let $n_i=|Q_{a_i}|$.
The proof is given separately for $n_1$ at most $\frac{n}{2}$ and for greater values of $n_1$.
\begin{itemize}
\item
Let $n_1 \leqslant \frac{n}{2}$.
The initial subset is always reachable: this is one subset.
Any other reachable subset is reached by some symbol $a_i$, with $1 \leqslant i \leqslant k$.
By reading this symbol, NFA $A$ can get only to states from $Q_{a_i}$,
and accordingly, this subset may contain only states from $Q_{a_i}$.
Hence, there are at most $2^{n_i}-1$ nonempty subsets reachable after reading $a_i$.
Also, the empty subset may be reachable: it is one subset for all symbols.
Therefore, the number of reachable subsets is bounded by $1 + 2^{n_1}+\ldots+2^{n_k}-k+1$.
It is claimed that this sum does not exceed $2^{\frac{n}{2}+1}$ in this case.
This will be proved by nested induction by $n_1$ downwards, with the inner induction by $n_2$ downwards.

The base case is when the values of $n_1$ and $n_2$ are maximum possible,
that is, $n_1=n_2=\lfloor\frac{n}{2}\rfloor$.
Then, in the case of even $n$,
all states are reachable by the first two symbols,
and the sum is
$1 + 2^{\frac{n}{2}} + 2^{\frac{n}{2}} - 1
= 2^{\frac{n}{2}+1}$.
And in the case of odd $n$, there must be the third symbol, with $n_3=1$ and $n_4=\ldots=n_k=0$,
and the sum accordingly is
$1 + 2^{\frac{n-1}{2}} + 2^{\frac{n-1}{2}} + 2^1 + (k-3) -k+1 = 2^{\frac{n+1}{2}} + 1 < 2^{\frac{n}{2}+1}$.

For the induction step, the first case
is when $n_1$ is not the maximum possible.
Then $n_1 + n_2 \leqslant 2n_1 < n$,
and hence there are non-zero values among $n_3, \ldots, n_k$.
Let $i \geqslant 3$ be the greatest index with $n_i>0$.
Then one can increase $n_1$ by $1$ and descrease $n_i$ by $1$:
the sum $n_1 + \ldots + n_k$ will remain equal to $n$,
and the inequalities $\frac{n}{2} \geqslant n_1 \geqslant \ldots \geqslant n_k$ will still hold.
The induction hypothesis is applicable to the modified vector of values
$(n_1+1, n_2, \ldots, n_{i-1}, n_i-1, n_{i+1}, \ldots, n_k)$,
and it asserts that the sum
$2^{n_1+1}+2^{n_i-1}+\sum_{j\neq 1,i} 2^{n_j}-k+2$ is bounded by $2^{\frac{n}{2}+1}$.
Since the desired sum is less by $2^{n_1}-2^{n_i-1} > 0$,
then it is also bounded.

The last case is when $n_1$ is the maximum possible,
whereas $n_2$ is not.
The proof is similar to the previous case:
the sum $2^{n_1}+\ldots+2^{n_k}-k+2$ can be increased
by incrementing $n_2$ and decrementing some $n_i$.
Then the induction hypothesis guarantees
that even the increased sum is at most $2^{\frac{n}{2}+1}$.

\item
Assume that the symbol $a_1$ occurs more often
than the rest of the symbols combined: $n_1 > \frac{n}{2}$.
The subset construction gives the following reachable subsets:
first, the set of initial states $Q_0$ (one subset),
secondly, some subsets of $Q_{a_1}$
and thirdly, some subsets of $Q_{a_2} \cup \ldots \cup Q_{a_k}$.
The number of subsets reachable by $a_1$
is at most $2^{n_1}$,
whereas the number of all other subsets is bounded by
$2^{n_2 + \ldots + n_k} + 1= 2^{n-n_1} + 1 < 2^{\lfloor\frac{n}{2}\rfloor} + 1$.
Overall, there are at most
$2^{n_1} + 2^{\lfloor\frac{n}{2}\rfloor} + 1 \leqslant 2^{n_1+1}$
subsets.
\qedhere
\end{itemize}
\end{proof}

If much more than $\frac{n}{2}$ states are reachable by some symbol $a \in \Sigma$,
then the subset construction can possibly yield many subsets reachable by this symbol.
In this case, the first method for subset calculation
gives at least $2^{|Q_a|}$ subsets.
However, the number of such subsets can be bounded by a different method,
which uses the known results on determinization of unary NFA.

\begin{figure}[t]
	\centerline{\includegraphics{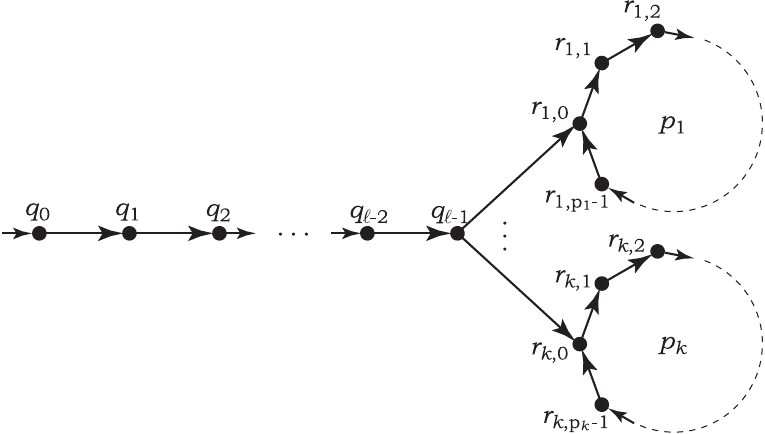}}
	\caption{The Chrobak normal form of unary NFA.}
	\label{f:chrobak_nf}
\end{figure}

Unary NFA can be determinized by first transforming them to the \emph{Chrobak normal form},
and then applying the subset construction, which, for automata in this form,
gives a predictable set of reachable subsets.

\begin{definition}[Chrobak~\cite{Chrobak}]
An NFA is said to be in the Chrobak normal form
if it is of the form $A=(\{a\}, Q, q_0, \delta, F)$,
where $Q=\{q_0, \ldots, q_{\ell-1}\} \cup \bigcup_{i=1}^k \{r_{i,0}, \ldots, r_{i,p_i-1}\}$,
for some $\ell \geqslant 1$, $k \geqslant 1$ and $p_1, \ldots, p_k \geqslant 1$,
and its transition function is as follows (see Figure~\ref{f:chrobak_nf}).
\begin{align*}
	\delta(q_i, a) &= q_{i+1}
		&& (0 \leqslant i < \ell-1) \\
	\delta(q_{\ell-1}, a) &= \{r_{1,0}, \ldots, r_{k,0}\}
		\\
	\delta(r_{i,j}, a) &= r_{i,j+1 \bmod{p_i}}
		&& (1 \leqslant i \leqslant k, \: 0 \leqslant j < p_i)
\end{align*}
The accepting states can be arbitrary.
\end{definition}

The subset construction applied to an NFA in the Chrobak normal form
produces a DFA
with a tail of length $\ell$
and one cycle of length $\lcm(p_1, \ldots, p_k)$.
The greatest possible length of this cycle for a given number of states in an NFA
is expressed through \emph{Landau's function}.
\begin{equation*}
	g(n)
		=
	\max\set{\lcm(p_1, \ldots, p_k)}{k \geqslant 1, \: p_1 + \ldots + p_k \leqslant n}
		=
	e^{(1+o(1))\sqrt{n \ln n}}
\end{equation*}

In this paper, for an NFA over an arbitrary alphabet,
the task is to estimate the number of subsets reachable by a string of the form $wa^t$,
where $t \geqslant 1$ and $w$ does not end with $a$,
and $a$ is the symbol with $|Q_a| > \frac{n}{2}$.
The idea is that the number of subsets reachable by $w$ will be bounded,
and starting from each of these subsets,
while reading $a^t$, the automaton works like a unary automaton.
Once any symbol other than $a$ is read,
the transitions from the current subset will lead back to the non-unary part,
that is, to some subset of $Q \setminus Q_a$.
Hence, all that is needed is to estimate the number of different subsets
reached in the unary automaton starting from any fixed subset.

If one simply transforms an NFA to the Chrobak normal form, and then determinizes it,
the result will only be a DFA recognizing the same language,
which may have fewer states
than the number of reachable subsets in the original NFA.
In particular, such a DFA would not know the exact subset reached by the original NFA at the current moment,
and hence would not be able to move on to the non-unary part upon reading a symbol other than $a$.

In order to estimate the number of reachable subsets,
a small extension of Chrobak's~\cite{Chrobak} construction will be proved.
It will transform an NFA to a DFA in such a way
that the states of the DFA are subsets \emph{of the original NFA}.

\begin{lemma}\label{extended_chrobak_lemma}
Let $A$ be a unary NFA without accepting states,
and let $C$ be a DFA made from $A$ by the subset construction.
Then $C$ has at most $g(n)+O(n^2)$ states reachable from the initial state.
\end{lemma}
\begin{proof}
The proof essentially uses the details of Chrobak's~\cite{Chrobak} construction.
For every state $q$ of the automaton $A=(\Sigma, Q, q_0, \delta)$,
consider the automaton $A_q=(\Sigma, Q, q_0, \delta, \{q\})$,
in which $q$ is the unique accepting state.
The automaton $A_q$ is transformed into the Chrobak normal form~\cite[Lemma~4.3]{Chrobak},
resulting in an NFA $B_q$ that recognizes the same language,
and has a tail of length at most $O(n^2)$ and cycles of length $p_1, \ldots, p_k$,
where $k$ is the number of strongly connected components in the automaton $A$,
and for each $i$, the number $p_i$
is the greatest common divisor of the lengths of all cycles
in the $i$-th strongly connected component.
It is important that the number of cycles and their lengths in each automaton $B_q$
depend only on the automaton $A$,
and do not depend on the choice of the accepting state $q$.
One can assume that tails of all automata $B_q$ are of the same length $\ell=O(n^2)$,
by rolling back all cycles whenever the original tail was shorter.
Then the automata $B_q$ differ from each other only in their accepting states.

Next, for each $q$, let $C_q$ be a DFA obtained from $B_q$ by the subset construction.
For different $q$, these automata also differ only in their accepting states,
and the size of each of them is $g(n)+\ell$.
Let $R$ be the set of states of each automaton $C_q$.

It is claimed that for every string $a^m$,
the set of states of the original NFA $A$
reached by the string $a^m$
contains exactly those states $q$,
for which the corresponding DFA $C_q$ accepts the string $a^m$.
Indeed, a state $q$ can be reached by the string $a^m$ in the automaton $A$
if and only if the automaton $A_q$ accepts the string $a^m$.
And this is equivalent to the acceptance of this string by automata $B_q$ and $C_q$,
because all of them recognize the same language.

Let $C$ be the DFA obtained from the NFA $A$ by the subset construction.
Now it will be proved that $C$ has at most $g(n)+\ell$ reachable states.
Let $S \subseteq Q$ be a reachable subset in $A$,
and let $a^m$ be the string by which it is reached from the initial subset.
Then, as proved above, $S=\set{q}{a^m \in L(C_q)}$.
Next, let $r \in R$ be the common state in all automata $C_q$,
in which these automata come after reading $a^m$
(this is the same state, because all $C_q$ differ only in their accepting states).
Then, $S=\set{q}{r \text{ is accepting in } C_q}$.
Therefore, the number of distinct subsets $S$ does not exceed the number of states $r \in R$---%
and there are exactly $g(n)+\ell$ such states.
\end{proof}

Now this construction will be used
to estimate the number of reachable subsets
in a given NFA over a non-unary alphabet, which remembers the last symbol.
And if $|Q_a|$ is large enough, then the new method will be more efficient than the first one.

Assume that the NFA is either in its initial subset,
or in a subset $S \subseteq Q_b$ reachable by some symbol $b \neq a$,
and assume that it makes a transition by $a$.
At this moment, it can potentially get
into any subset of the unary part.
However, the number of such transitions is bounded by the number of subsets reachable not by $a$---%
and there are few of them by the assumption.
Thus, we have to deal with a bounded number of starting subsets of the set $Q_a$,
and for each of them one can work with a separate unary automaton
that starts from the given subset of $Q_a$,
and then reads symbols $a$, moving through different subsets of $Q_a$.

\begin{lemma}[The second method for subset calculation]
Let $A=(\Sigma, Q, Q_0, \delta, F)$
be an NFA with $n$ states that remembers the last symbol.
For each symbol $a \in \Sigma$,
let $Q_a \subseteq Q$ be all states reachable by $a$.
Denote the maximal number of states reachable by the same symbol
by $n_1 = \max_{a \in \Sigma} |Q_a|$.
Then there is a DFA with at most
$2^{n-n_1} \cdot (g(n_1)+O(n_1^2))$ states
recognizing the same language.
\end{lemma}
\begin{proof}
Let $\Sigma = \{a_1,\ldots, a_k\}$,
and let $|Q_{a_1}| \geqslant \ldots \geqslant |Q_{a_k}| \geqslant 0$.
Then $n_1=|Q_{a_1}|$.

The subset construction gives the following reachable subsets:
the set of initial states $Q_0$,
some subsets of $Q_{a_1}$,
and some subsets of $Q_{a_2} \cup \ldots \cup Q_{a_k}$.
At most $2^{n_1}$ subsets are reachable by $a_1$,
whereas the number of other subsets is bounded by $2^{n-n_1} + 1$.

The task is to improve the trivial $2^{n_1}$ bound
on the number of subsets of $Q_{a_1}$ in the NFA $A$
that are reachable by some string.
Every string, by which the NFA gets to states in $Q_{a_1}$,
is of the form $wa_1^t$, with $t \geqslant 1$,
and with the string $w$ either empty
or ending with a symbol different from $a_1$.
Let $P_w$ be the set of states of the NFA
reachable after reading $w$.
If $w$ is empty, then $P_w=Q_0$,
and if $w$ ends with some symbol $a_i$, with $i = 2,\ldots,k$,
then $P_w \subseteq Q_{a_i}$.
These are all starting subsets $P_w$ before the unary part of the string,
and the number of such subsets is at most $2^{n-n_1} + 1$.
It remains to estimate how many different subsets
are obtained from each starting subset
by reading suffixes from $a_1^*$.
This is done by a unary NFA with $n_1$ states,
which is a part of the original NFA.
Then, Lemma~\ref{extended_chrobak_lemma} is applicable to this unary NFA,
and it asserts that the subset construction,
with the given starting subset as initial,
gives at most $g(n_1)+O(n_1^2)$ reachable subsets.

Therefore, the overall number of reachable subsets in the constructed DFA
does not exceed $(2^{n-n_1} + 1)(g(n_1)+O(n_1^2))$.
\end{proof}

The minimum of the bounds given by the two methods
is an upper bound on the size of a DFA.
If $n_1$ is small, then the first method is better,
and if it is large, then the second one.
The desired minimum is estimated in the following theorem.

\begin{theorem}\label{NFA_remembering_last_symbol_to_DFA_upper_bound}
Let $A$ be an NFA that remembers the last symbol,
and let $n$ be the number of states in it.
Then there exists a DFA with at most
$2^{\frac{n}{2} + \frac{\log_2 e}{2 \sqrt{2}}\sqrt{n \ln n}(1+o(1))}$
states recognizing the same language.
\end{theorem}
\begin{proof}
Let $a$ be the symbol, by which the greatest number of states are accessed,
and $n_1$ be the number of states accessed by $a$.
If $n_1 < \frac{n}{2}$,
then the first method gives an upper bound of
$\max(2^{\frac{n}{2}+1}, 2^{n_1+1})=2^{\frac{n}{2}+1}$ states,
and this is less than the stated upper bound.
Hence, it is sufficient to consider the case of $n_1 \geqslant \frac{n}{2}$,
for which the first method gives an upper bound of $2^{n_1+1}$ states.

For each $n$, the goal is to find the worst-case value of $n_1$,
for which the minimum over the two methods will be the greatest.
The upper bound given by the first method, $2^{n_1+1}$, grows with $n_1$.
At the first glance, the bound given by the second method, $2^{n-n_1} \cdot (g(n_1)+O(n_1^2))$,
should be decreasing with $n_1$ for $n$ large enough,
because Landau's function grows asymptotically slower than the exponential function.
However, for some values of $n_1$,
the value of Landau's function may double
with the argument increasing by one.
At such points, the function $2^{n-n_1} \cdot (g(n_1)+O(n_1^2))$ will marginally increase.
Hence, it would be difficult to obtain the optimal value of $n_1$ by equating the two functions.

The plan is to prove the estimation as follows.
First, the upper bound from the second method
will be replaced with an asymptotically close descreasing continuous function.
This function is then equated to the bound from the first method,
and a solution of the resulting equation will be found.
Let $n_1=x=x(n)$ be this solution.
Up to this point, the value of $x$ is nothing more than a heuristic.
Then, a strict proof is done
by substituting this $x$ into the bounds from both methods,
and verifying that for values $n_1 \leqslant x$,
the first method gives no more than the desired upper bound stated in the theorem,
whereas for values $n_1 \geqslant x$,
the second method gives at most the stated bound.
Furthermore, it will turn out that for $n_1=\lfloor x \rfloor$
both methods give the asymptotic upper bound from the statement of the theorem,
and therefore no better upper bound can be obtained using these two methods.

Using the asymptotics of Landau's function $g(n)=e^{(1+o(1))\sqrt{n \ln n}}$,
the upper bound given by the second method
can be approximated as follows.
\begin{equation*}
	2^{n-n_1} \cdot (g(n_1)+O(n_1^2))
		=
	2^{n-n_1} e^{(1+o(1))\sqrt{n_1 \ln n_1}}
		=
	2^{n-n_1 + (1+o(1))(\log_2 e) \sqrt{n_1 \ln n_1}}
\end{equation*}
This bound is equated to the bound given by the first method.
\begin{equation*}
	2^{n-n_1 + (1+o(1))(\log_2 e) \sqrt{n_1 \ln n_1}} = 2^{n_1+1}
\end{equation*}
Then the exponents are equal.
\begin{equation*}
	n-n_1 + (1+o(1))(\log_2 e) \sqrt{n_1 \ln n_1} = n_1
\end{equation*}
The optimal value of $n_1$, for which this equation will hold,
will be searched for in the form of $n_1=\frac{n}{2} + c\sqrt{n \ln n}$.
A substitution into both the left-hand side and the right-hand side gives
\begin{equation*}
	\frac{n}{2} - c\sqrt{n \ln n}
	+
	(1+o(1))(\log_2 e)
	\sqrt{\big(\frac{n}{2} + c\sqrt{n \ln n}\big) \ln \big(\frac{n}{2} + c\sqrt{n \ln n}\big)}
		=
	\frac{n}{2} + c\sqrt{n \ln n}
\end{equation*}
The square root in this equation is estimated like this.
\begin{equation*}
	\sqrt{\big(\frac{n}{2} + c\sqrt{n \ln n}\big) \ln \big(\frac{n}{2} + c\sqrt{n \ln n}\big)}
		=
	(1+o(1))\frac{1}{\sqrt{2}}\sqrt{n \ln n}
\end{equation*}
Then the equation becomes
\begin{equation*}
	\frac{n}{2} - c\sqrt{n \ln n}
	+
	(1+o(1))
	\frac{\log_2 e}{\sqrt{2}}
	\sqrt{n \ln n}
		=
	\frac{n}{2} + c\sqrt{n \ln n}
\end{equation*}
After combining like terms:
\begin{equation*}
	(1+o(1))
	\frac{\log_2 e}{\sqrt{2}}
	\sqrt{n \ln n}
		=
	2c\sqrt{n \ln n}
\end{equation*}
Therefore,
\begin{equation*}
	c
		=
	(1+o(1))
	\frac{\log_2 e}{2\sqrt{2}}
\end{equation*}
Accordingly, the informally defined point of equilibrium for the parameter $n_1$
has been found in the form
\begin{equation*}
	x
		=
	\frac{n}{2}
	+
	(1+o(1))
	\frac{\log_2 e}{2\sqrt{2}}
	\sqrt{n \ln n}
\end{equation*}

Now, using this value, the theorem will be proved formally.
Consider two cases.
The first case: $n_1 \leqslant x$.
Then the bound of $2^{n_1 + 1}$ states given by the first method
can be estimated from above like this.
\begin{equation*}
	2^{n_1 + 1}
		\leqslant
	2^{x+1}
		=
	2^{%
	\frac{n}{2}
	+
	(1+o(1))
	\frac{\log_2 e}{2\sqrt{2}}
	\sqrt{n \ln n}
	}
\end{equation*}
This is the desired upper bound.

The second case: $n_1 \geqslant x$.
Here the number of states is estimated by the second method as
\begin{align*}
	2^{n-n_1} \cdot (g(n_1)+O(n_1^2))
		=
	2^{n-n_1 + (1+f(n_1))(\log_2 e) \sqrt{n_1 \ln n_1}},
	&& \text{where } f(n_1)=o(1)
\end{align*}
Since $\frac{n}{2} < n_1 \leqslant n$,
the function $\widehat{f}(n)=\max_{\frac{n}{2} < n_1 \leqslant n} f(n_1)$
is also infinitesimal
and depends on $n$ rather than on $n_1$.
Next, let $h(n)=\max_{t=n}^\infty \widehat{f}(t)$.
Then $f(n_1) \leqslant \widehat{f}(n) \leqslant h(n)$,
and $h(n)$ is descreasing and infinitesimal.
Then, in the upper bound on the number of states
one can replace $f(n_1)$ with $h(n)$,
and then replace $n_1$ with $x$.
This is possible, because $n_1 > x > \frac{n}{2}$,
and the contribution of the linear term $-n_1$
exceeds the contribution of the square root $\sqrt{n_1 \ln n_1}$
multiplied by a constant-bounded factor that is not affected
by the substitution of $x$ for $n_1$.
\begin{align*}
	2^{n-n_1 + (1+f(n_1))(\log_2 e) \sqrt{n_1 \ln n_1}}
	\leqslant
	2^{n-n_1 + (1+h(n))(\log_2 e) \sqrt{n_1 \ln n_1}}
	\leqslant \\ \leqslant
	2^{n-x + (1+h(n))(\log_2 e) \sqrt{x \ln x}}
\end{align*}
Now, after the elimination of $n_1$, the value of $x$ can be substituted into the square root.
\begin{multline*}
	\sqrt{x \ln x}
		=
	\sqrt{%
	\Big(
	\frac{n}{2}
	+
	(1+o(1))
	\frac{\log_2 e}{2\sqrt{2}}
	\sqrt{n \ln n}
	\Big)
	\ln\Big(
	\frac{n}{2}
	+
	(1+o(1))
	\frac{\log_2 e}{2\sqrt{2}}
	\sqrt{n \ln n}
	\Big)
	}
		= \\ =
	(1+o(1))
	\frac{1}{\sqrt{2}}
	\sqrt{n \ln n}
\end{multline*}
Finally, the values of $x$ and $\sqrt{x \ln x}$
are substituted into the whole expression for the upper bound.
\begin{multline*}
	2^{n-x + (1+o(1))(\log_2 e) \sqrt{x \ln x}}
		=
	2^{n-\frac{n}{2}
	- (1+o(1)) \frac{\log_2 e}{2\sqrt{2}} \sqrt{n \ln n}
	+ (1+o(1))(\log_2 e)
	\frac{1}{\sqrt{2}}
	\sqrt{n \ln n}
	}
		= \\ =
	2^{\frac{n}{2}
	+ (1+o(1)) \frac{\log_2 e}{2\sqrt{2}} \sqrt{n \ln n}
	}
\end{multline*}
The desired upper bound has thus been established in both cases.
Furthermore, under a substitution of $\lfloor x \rfloor$ for $n_1$
both calculations resulted in functions of the order
$2^{\frac{n}{2}
	+ (1+o(1)) \frac{\log_2 e}{2\sqrt{2}} \sqrt{n \ln n}
	}$.
Therefore, the minimum over the two calculation methods
cannot yield a better upper bound than this function.
\end{proof}
 
\begin{corollary}\label{Reg_exp_to_DFA_upper_bound_corollary}
Let $\alpha$ be a regular expression of size $n$ over an alphabet $\Sigma$.
Then there is a DFA with at most
$2^{\frac{n}{2} + \frac{\log_2 e}{2 \sqrt{2}}\sqrt{n \ln n}(1+o(1))}$
states that recognizes the same language.
\end{corollary}

\section{Lower bound}\label{section_lower_bound}

In this section, a lower bound on the number of states in a DFA
necessary to recognize languages
that are defined by regular expressions of a given size
is obtained.
Also, the same lower bound will be shown
for the number of states necessary to determinize NFA that remember the last symbol.

\begin{theorem}\label{NFA_remembering_last_symbol_to_DFA_lower_bound_theorem}
For every $n \geqslant 2$ there exists a language
defined by a regular expression of size $n$,
and at the same time recognized by an NFA that remembers the last symbol, with $n-1$ states,
such that every DFA recognizing this language
has at least
$2^{\frac{n}{2} + (1+o(1)) \sqrt{2}\sqrt{\frac{n}{\ln n}}}$
states.
\end{theorem}

The general form of regular expressions used in the proof of the theorem is as follows.

\begin{definition}
Let $\pi_1, \ldots, \pi_k \geqslant 3$ be relatively prime numbers, with $\pi_1=3$.
For this set of numbers, a regular expression $\alpha_{\pi_1, \ldots, \pi_k}$ of the following form is defined.
\begin{align*}
	\alpha_{\pi_1, \ldots, \pi_k} &= \big(a(\beta_{\pi_1} \ | \ \ldots \ | \ \beta_{\pi_k})b\big)^*,
		&& \text{where }
		\beta_\pi=(a((b\:|\:\epsilon)a)^{\pi-2}a)^*.
\end{align*}
The size of every subexpression $\beta_\pi$ is $2\pi-2$.
Thus, the size of $\alpha_{\pi_1, \ldots, \pi_k}$
is $2+\sum_{i=1}^k (2\pi_i-2) = (2\sum_{i=1}^k \pi_i) - 2k + 2$.
\end{definition}

It will be shown that, with a proper choice of numbers $\pi_1, \ldots, \pi_k$
satisfying the inequality $(2\sum_{i=1}^k \pi_i) - 2k + 2 \leqslant n$,
the regular expression $\alpha_{\pi_1, \ldots, \pi_k}$ will be the desired example for the theorem.
The numbers $\pi_1, \ldots, \pi_k$ will be chosen later.

To prove that a DFA for this language needs a lot of states,
it would be easier to consider an NFA recognizing the same language.
The suggested NFA slightly differs from the automaton
given by the formal transformation from Lemma~\ref{regular_expression_to_nfa_lemma};
in particular, it has two less states.

\begin{figure}[t]
\centerline{\includegraphics[scale=1]{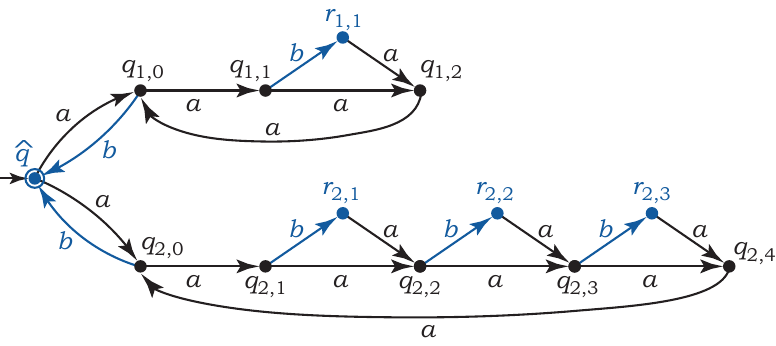}}
\caption{NFA $A_{3,5}$.}
\label{f:nfa_3_5}
\end{figure}

\begin{definition}
An NFA $A_{\pi_1, \ldots, \pi_k}=(\{a,b\}, Q, \widehat{q}, \delta, F)$
has the set of states
$Q=\{\widehat{q}\} \cup \bigcup_{i=1}^k Q_i \cup \bigcup_{i=1}^k R_i$,
where $Q_i=\{q_{i,0}, \ldots, q_{i,\pi_i-1}\}$
and $R_i=\{r_{i,1}, \ldots, r_{i,\pi_i-2}\}$.

Its only transition in the initial state is
\begin{align*}
	\delta(\widehat{q}, a) &= \{q_{1,0}, \ldots, q_{k,0}\}
\intertext{%
and this is the only nondeterministic transition in the automaton.
Each subset $Q_i$ has a cycle by $a$ of length $\pi_i$, and is called \emph{the $i$-th cycle}:
}
	\delta(q_{i,j},a) &= q_{i,j+1 \bmod \pi_i},
		&& \text{for all } j.
\intertext{%
For each state in $Q_i$, except the first and the last ones,
there is a transition by $b$ to a separate state from $R_i$:
}
	\delta(q_{i,j},b) &= r_{i,j},
		&& \text{with } 1 \leqslant j \leqslant \pi_i-2.
\intertext{%
From every such state the automaton moves by $a$ to the next state in $Q_i$:
}
	\delta(r_{i,j},a) &=q_{i,j+1},
		&& \text{with } 1 \leqslant j \leqslant \pi_i-2.
\intertext{%
Finally, from the first state in each $Q_i$ there is a transition by $b$ to the initial state:
}
	\delta(q_{i,0},b) &= \widehat{q}.
\end{align*}
The only accepting state is the initial state: $F=\{\widehat{q}\}$.
In each set $Q_i \cup R_i$ there are $2\pi_i-2$ states in total.
Hence, the overall number of states in the automaton
is $2(\sum_{i=1}^k \pi_i) - 2k + 1$.

The automaton $A_{3,5}$ is shown in Figure~\ref{f:nfa_3_5}.
\end{definition}

\begin{lemma}
Let $\pi_1, \ldots, \pi_k \geqslant 3$ be any relatively prime integers, with $\pi_1=3$.
Then the automaton $A_{\pi_1, \ldots, \pi_k}$
and the regular expression $\alpha_{\pi_1, \ldots, \pi_k}$
define the same language.
\end{lemma}
\begin{proof}[Sketch of a proof]
Each regular expression $\beta_{\pi_i}$ defines the set of all strings,
on which the automaton, having started in the state $q_{i,0}$,
visits only states from $Q_i \cup R_i$,
and finally returns to the state $q_{i,0}$.
Therefore, both the regular expression $\alpha_{\pi_1, \ldots, \pi_k}$
and the automaton $A_{\pi_1, \ldots, \pi_k}$
define the language of all strings of the form $ax_1bax_2b \ldots ax_\ell b$,
where every segment $x_t$ is a substring defined by the regular expression $\beta_{\pi_{i_t}}$,
for some $i_t$.
\end{proof}

\begin{lemma}\label{state_complexity_lower_bound_lemma}
Let $k \geqslant 2$,
and let $\pi_1, \ldots, \pi_k \geqslant 3$ be any relatively prime integers, where $\pi_1=3$.
Then the minimal DFA that recognizes the same language as the NFA $A_{\pi_1, \ldots, \pi_k}$,
has at least $\prod_{i=1}^k (2^{\pi_i}-2)$ states.
\end{lemma}
\begin{proof}
The subset construction is applied to $A_{\pi_1, \ldots, \pi_k}$.

The proof uses subsets of the form $P_1 \cup \ldots \cup P_k$, where $P_i \subseteq Q_i$,
in which, from each $Q_i$, at least one state is taken and at least one state is not taken:
$\emptyset \subset P_i \subset Q_i$.
There are $2^{\pi_i}-2$ possible choices for each $P_i$,
and hence there are $\prod_{i=1}^k (2^{\pi_i}-2)$ subsets of this form.

To subsets of such a form, one can conveniently apply
an operation of \emph{shift by a vector of residues} $(d_1, \ldots, d_k)$,
where $d_i \in \{0, \ldots, \pi_i-1\}$:
this means reading a substring $a^\ell$, where $\ell \equiv d_i \pmod{\pi_i}$ for all $i$.
Since the cycle lengths $\pi_1, \ldots, \pi_k$ are relatively prime,
such a number $\ell$ exists by the Chinese remainder theorem.
This operation transforms the subset $P_1 \cup \ldots \cup P_k$
into the subset $P'_1 \cup \ldots \cup P'_k$,
where $P'_i=\set{q_{i,j+d_i \bmod{\pi_i}}}{q_{i,j} \in P_i}$ for all $i$.

The overall plan of the proof is standard:
first, the reachability of all subsets of the chosen form is established,
and then a separating string is found for every pair of distinct subsets of this form.
Once this is done, this will imply
that every DFA recognizing this language
must have at least as many states as there are subsets of this form.

\begin{claim}\label{state_complexity_lower_bound_lemma__reachability_claim}
Each subset of the form $P_1 \cup \ldots \cup P_k$,
with $P_i \subseteq Q_i$ and $\emptyset \subset P_i \subset Q_i$,
is reachable from the initial subset $\{\widehat{q}\}$
by some string.
\end{claim}
Induction on the number of states in a subset.

Base case: $|P_1| \in \{1, 2\}$ and $|P_2|=\ldots=|P_k|=1$.
If $|P_1|=1$, then the subset
$\{q_{1,0}\} \cup \{q_{2,0}\} \cup \ldots \cup \{q_{k,0}\}$
is reached by a single transition by the symbol $a$,
and then the desired subset is reached by applying a shift
by the vector of residues $(d_1, \ldots, d_k)$, where $P_i=\{q_{i,d_i}\}$ for all $i$.
If $|P_1|=2$,
then the subset
$\{q_{1,0}, q_{1,2}\} \cup \{q_{2,0}\} \cup \ldots \cup \{q_{k,0}\}$
is reached by first moving by $a$,
then applying a shift by the vector $(1, 0, \ldots, 0)$,
and then reading $ba$.
The rest of the subsets in the second part of the base case
are reached from this one by applying another shift.

Induction step.
Let $P_1 \cup \ldots \cup P_k$ be any subset
satisfying the condition and not handled in the base case.
Then $|P_m|>1$ for some $m \geqslant 2$.
Let this $m$ be fixed.
In each subset $P_i$, a state is chosen as follows.
In the first cycle, the chosen state is the unique state $q_{1,s}$,
such that $q_{1,s} \in P_1$ and $q_{1,s-1 \bmod{3}} \notin P_1$.
In each of the subsequent cycles $Q_i$, with $i \in \{2, \ldots, k\}$,
any state $q_{i,t_i}$ in $P_i$ can be chosen,
for which the next state is not in $P_i$,
that is, $q_{i,t_i+1 \bmod{\pi_i}} \notin P_i$.
Then the set $(P'_1, \ldots, P'_k)$,
where $P'_m=P_m \setminus \{q_{m,t_m}\}$
and $P'_i=P_i$ for all $i \neq m$,
satisfies the condition
and contains fewer states that the desired subset.
Therefore, it is reachable by the induction hypothesis.

It remains to reach the subset $(P_1, \ldots, P_k)$
starting from the subset $(P'_1, \ldots, P'_k)$.
To this end, the subset $(P'_1, \ldots, P'_k)$
is first shifted by the vector of residues $(-s, -t_2-1, \ldots, -t_k-1)$,
resulting in the subset
$\set{q_{1,j-s \bmod{\pi_1}}}{q_{i,j} \in P_1}
	\cup
\set{q_{m,j-t_m-1 \bmod{\pi_i}}}{q_{m,j} \in P_m \setminus \{q_{m,t_m}\}}
	\cup
\bigcup_{i \geqslant 2, \: i \neq m}^k \set{q_{i,j-t_i-1 \bmod{\pi_i}}}{q_{i,j} \in P_i}$.
In particular, the first cycle will contain the state $q_{1,0}$, but not the state $q_{1,2}$;
in the $m$-th cycle there will be neither $q_{m,0}$ nor $q_{m,\pi_m-1}$;
and in the rest of the cycles, for all $i \notin \{0, m\}$,
there will be no $q_{i,0}$, but $q_{i,\pi_i-1}$ will be present.

\begin{figure}[t]
\centerline{\includegraphics[scale=1]{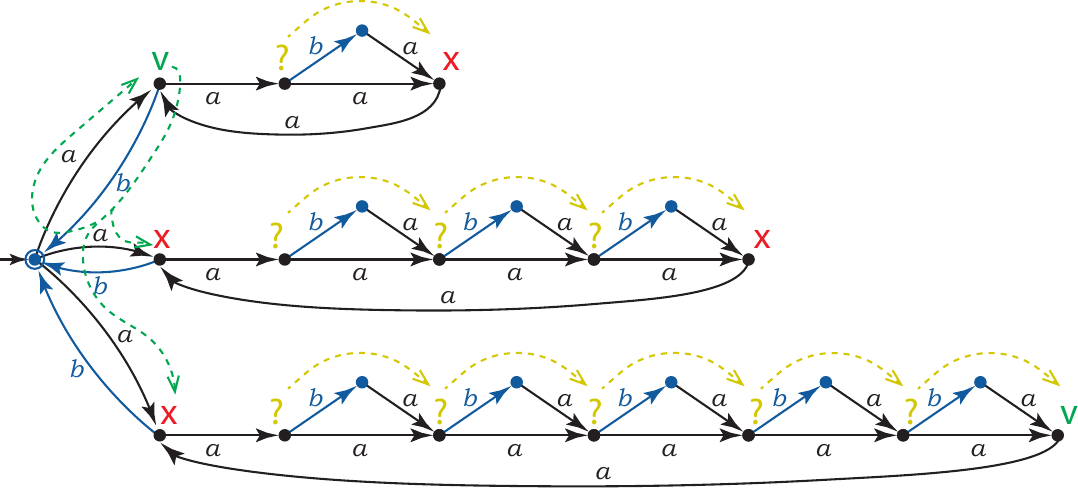}}
\caption{Proof of Claim~\ref{state_complexity_lower_bound_lemma__reachability_claim}
	in Lemma~\ref{state_complexity_lower_bound_lemma}:
	for NFA $A_{3,5,7}$,
	applying the string $ba$ to add a a state to the cycle $P_m$, where $m=2$.}
\label{f:nfa_3_5_7_reachability}
\end{figure}

Then the string $ba$ is read;
its effect on the states is illustarted in Figure~\ref{f:nfa_3_5_7_reachability}.
It affects the first cycle as follows:
if it contained only the state $q_{1,0}$,
then this state will move to $\widehat{q}$ by $b$ and return to $q_{1,0}$ by $a$.
And if there were both states $q_{1,0}$ and $q_{1,1}$,
then they move to $q_{1,0}$ and the $q_{1,2}$, respectively;
that is, the cycle moves forward by two positions,
and the resulting states are $\set{q_{1,j-s+2 \bmod{3}}}{q_{i,j} \in P_1}$.

In the $m$-th cycle, after reading $ba$, the following happens.
All states $q_{m,1}, \ldots, q_{m,\pi_m-2}$ move by one position forward.
After reading the first symbol $b$, the state $\widehat{q}$ emerges from the first cycle,
and after reading $a$ it gets copied to $q_{m,0}$.
As the result, the states in the $m$-th cycle will be
$\set{q_{m,j-t_m \bmod{\pi_m}}}{q_{m,j} \in P_m \setminus \{q_{m,t_m}\}} \cup \{q_{m,0}\} =
\set{q_{m,j-t_m \bmod{\pi_m}}}{q_{m,j} \in P_m}$.

In every $i$-th cycle, with $i \notin \{1, m\}$,
by reading $ba$,
the states $q_{i,1}, \ldots, q_{i,\pi_i-2}$ are also shifted forward by one position,
while the state $q_{i,\pi_i-1}$ disappears after $b$
but gets restored into $q_{i,0}$ upon reading $a$ from $\widehat{q}$.
Thus, the entire $i$-cycle is shifted forward by one position,
and it will contain the states $\set{q_{i,j-t_i \bmod{\pi_i}}}{q_{i,j} \in P_i}$.

In order to reach the desired subset $P_1 \cup \ldots \cup P_k$,
it is enough to shift the subset obtained after reading $ba$
either by the vector $(s, t_1, \ldots, t_k)$, if $|P_1|=1$,
or by $(s-2, t_1, \ldots, t_k)$, if $|P_1|=2$.
This proves the induction step
and the entire Claim~\ref{state_complexity_lower_bound_lemma__reachability_claim}.

\begin{claim}
For every two distinct subsets of the form $P_1 \cup \ldots \cup P_k$,
with $P_i \subseteq Q_i$ and $\emptyset \subset P_i \subset Q_i$,
there is a string that is accepted from exactly one of these subsets.
\end{claim}
Let $P_1 \cup \ldots \cup P_k$ and $P'_1 \cup \ldots \cup P'_k$
be two such subsets,
and let $q_{i,j}$ be the state on which they differ.
Assume, without loss of generality, that $q_{i,j} \in P_i \setminus P'_i$.
For each $t \neq i$, let $j_t$ be such a residue modulo $\pi_i$,
that $q_{t,j_t} \notin P'_i$.
By Chinese remainder theorem, there is such a length $\ell \geqslant 0$,
that after reading $a^\ell$ the state $q_{i,j}$ moves to $q_{i,0}$,
while every state $q_{t,j_t}$ moves to $q_{t,0}$.
Then the desired separating string is $a^\ell b$.
After reading this string from the subset $P_1 \cup \ldots \cup P_k$,
the automaton will accept,
since the state $q_{i,j}$ moves to $q_{i,0}$ by $a^\ell$,
and then by $b$ it moves to the accepting state $\widehat{q}$.
But the automaton does not accept this string
starting from the subset $P'_1 \cup \ldots \cup P'_k$,
because after reading $a^\ell$,
in each $t$-th cycle, for $1 \leqslant t \leqslant k$,
there will be no state $q_{t,0}$,
and hence, none of the states can move to $\widehat{q}$ after reading $b$.

With the above two claims established, the lemma is proved.
\end{proof}

In the lower bound expression in Lemma~\ref{state_complexity_lower_bound_lemma},
inconvenient summands $-2$ in all factors
can be eliminated at the cost of a constant factor.

\begin{lemma}
Let $\pi_1, \ldots, \pi_k \geqslant 3$ be any distinct integers.
Then $\prod_{i=1}^k (2^{\pi_i}-2) \geqslant \frac{1}{2} \prod_{i=1}^k 2^{\pi_i}$.
\end{lemma}
\begin{proof}
Without loss of generality, let $\pi_1 < \ldots < \pi_k$.
First, the product is transformed as follows.
\begin{equation*}
\prod_{i=1}^k (2^{\pi_i}-2)
	=
\prod_{i=1}^k \Big(1-\frac{2}{2^{\pi_i}}\Big) 2^{\pi_i}
	=
\bigg(\prod_{i=1}^k \Big(1-\frac{2}{2^{\pi_i}}\Big)\bigg) \cdot \bigg(\prod_{i=1}^k 2^{\pi_i}\bigg)
\end{equation*}
The first factor will be estimated using the following general inequality.
\begin{claim}
For any real numbers $x_1, \ldots, x_k \in (0, 1)$,
the inequality $\prod_{i=1}^k (1-x_i) \geqslant 1 - \sum_{i=1}^k x_i$ holds.
\end{claim}
\begin{proof}
Induction on $k$.
Base case $k=1$: true.

Induction step. Let $k \geqslant 2$ and
$\prod_{i=1}^{k-1} (1-x_i) \geqslant 1 - \sum_{i=1}^{k-1} x_i$.
Then
\begin{equation*}
	\prod_{i=1}^k (1-x_i)
		=
	(1 - x_k)\prod_{i=1}^{k-1} (1-x_i)
		\geqslant
	(1 - x_k)\bigg(1 - \sum_{i=1}^{k-1} x_i\bigg)
		=
	1 - x_k - \sum_{i=1}^{k-1} x_i + x_k \sum_{i=1}^{k-1} x_i
		\geqslant
	1 - \sum_{i=1}^k x_i
\end{equation*}
\end{proof}
Using this inequality and the fact that $\pi_i \geqslant i+2$ for all $i$,
\begin{equation*}
	\bigg(\prod_{i=1}^k \Big(1-\frac{2}{2^{\pi_i}}\Big)\bigg) \cdot \bigg(\prod_{i=1}^k 2^{\pi_i}\bigg)
		\geqslant
	\Big(1 - \sum_{i=1}^k \frac{2}{2^{\pi_i}} \Big) \cdot \prod_{i=1}^k 2^{\pi_i}
		\geqslant
	\Big(1 - \sum_{i=1}^\infty \frac{2}{2^{i+2}} \Big) \cdot \prod_{i=1}^k 2^{\pi_i}
		=
	\frac{1}{2} \cdot \prod_{i=1}^k 2^{\pi_i}.
\end{equation*}
\end{proof}

It remains to choose the optimal cycle lengths to obtain the desired lower bound for each $n$.

Let $n \geqslant 5$, and assume that $k$ cycles of length $\pi_1, \ldots, \pi_k$ are used.
Then the size of the regular expression is $S=2\sum_{i=1}^k \pi_i - 2k + 2$,
and it should be not greater than $n$.
Then $\sum_{i=1}^k \pi_i = \frac{S}{2}+k-1$,
and lower bound obtained can be represented as
\begin{equation*}
	\frac{1}{2} \cdot \prod_{i=1}^k 2^{\pi_i}
		=
	\frac{1}{2} \cdot 2^{\sum_{i=1}^k \pi_i}
		=
	\frac{1}{2} \cdot 2^{\frac{S}{2}+k-1}
		=
	2^{\frac{S}{2}+k-2}.
\end{equation*}
If the size $S$ of the regular expression is equal or almost equal to $n$,
then the resulting lower bound is of the order $\Omega(2^{\frac{n}{2} + k})$,
and hence in this case one should use as many cycles as possible,
that is, maximize $k$.

Simply maximizing the number of cycles is easy:
it is sufficient to use as many first odd primes as possible for cycle lengths.
Let $p_i$ be the $i$-th odd prime, that is, $p_1=3$, $p_2=5$, \ldots,
and let $k=k(n)$ be the greatest such number that
$S=(2\sum_{i=1}^k p_i) - 2k + 2 \leqslant n$.
However, in this case $S$ may be substantially less than $n$,
possibly by as much as $2p_{k+1}-1$.
How much is that?
To see this, one should express $k$ through $S$ and then estimate $p_{k+1}$.

The well-known asymptotic formula for the sum of the first $k$ primes is $(1+o(1)) \frac{k^2 \ln k}{2}$.
Hence, the sum of the first $k$ odd primes has the same asymptotics:
$\sum_{i=1}^k p_i = (1+o(1)) \frac{k^2 \ln k}{2}$.

The regular expression $\alpha_{p_1, \ldots, p_k}$ made for the first $k$ odd primes
is of size $S=2\sum_{i=1}^k p_i - 2k + 2 = (1+o(1)) k^2 \ln k$.
In order to estimate the asymptotic dependence of $k$ on $S$,
it is convenient to move to continuous functions.

\begin{lemma}\label{sum_of_k_primes_inversion_lemma}
Let $f \colon \mathbb{R} \to \mathbb{R}$ be a function
that is continuous and increasing on the interval $[1, \infty)$,
and let $f(x) \sim x^2 \ln x$ as $x \to \infty$.
Then $f^{-1}(y) \sim \sqrt{\frac{2y}{\ln y}}$ as $y \to \infty$.
\end{lemma}
\begin{proof}
The lemma is proved like a similar claim in Miller's~\cite[\S 6]{Miller} paper.

The proof is by contradiction: suppose that $f^{-1}(y) \not\sim \sqrt{\frac{2y}{\ln y}}$.
This means that there are arbitrarily large values of $y$
on which $f^{-1}(y)$ is not in the vicinity of $\sqrt{\frac{2y}{\ln y}}$.
There are two cases: the first case is when there exists $\epsilon > 0$,
for which
\begin{equation*}
	f^{-1}(y) \leqslant (1-\epsilon)\sqrt{\frac{2y}{\ln y}}
\end{equation*}
for $y$ arbitrarily large.
Then, since the function $\varphi(x)=x^2 \ln x$ is increasing,
it can be applied to both sides of the latter inequality,
resulting in the next inequality,
which accordingly also holds for arbitrarily large values of $y$.
\begin{equation*}
	\varphi(f^{-1}(y))
		\leqslant
	(1-\epsilon)^2 \frac{2y}{\ln y}
	\Big(\ln(\sqrt{2}(1-\epsilon)) + \frac{1}{2}\ln y - \frac{1}{2} \ln\ln y\Big)
\end{equation*}
Division by $y$ gives an equivalent inequality,
in which $\frac{2}{\ln y}$ gets cancelled in the right-hand side.
\begin{multline*}
	\frac{\varphi(f^{-1}(y))}{y}
		\leqslant
	(1-\epsilon)^2 \frac{2}{\ln y}
	\Big(\ln(\sqrt{2}(1-\epsilon)) + \frac{1}{2}\ln y - \frac{1}{2} \ln\ln y\Big)
		= \\ =
	(1-\epsilon)^2
	\Big(\frac{2}{\ln y}\ln(\sqrt{2}(1-\epsilon)) + 1 - \frac{\ln\ln y}{\ln y}\Big)
\end{multline*}
It turns out that the right-hand side tends to $(1 - \epsilon)^2$ as $y \to \infty$.
And what does the left-hand side tend to?
Since $\varphi(x) \sim f(x)$ as $x \to \infty$,
whereas $f^{-1}(y) \to \infty$ as $y \to \infty$,
it follows that $\varphi(f^{-1}(y)) \sim f(f^{-1}(y)) = y$.
This means that the left-hand side tends to 1 as $y \to \infty$,
and therefore the inequality can hold only for bounded values of $y$.

Now consider the second case, when, for some $\epsilon > 0$,
the inequality
\begin{equation*}
	f^{-1}(y) \geqslant (1+\epsilon)\sqrt{\frac{2y}{\ln y}}
\end{equation*}
holds for arbitrary large $y$.
Applying $\varphi$ to both sides gives
\begin{equation*}
	\varphi(f^{-1}(y))
		\geqslant
	(1+\epsilon)^2 \frac{2y}{\ln y}
	\Big(\ln(\sqrt{2}(1+\epsilon)) + \frac{1}{2}\ln y - \frac{1}{2} \ln\ln y\Big)
\end{equation*}
Similarly to the previous case,
after division by $y$ the left-hand side will tend to 1,
while the right-hand side will tend to $(1+\epsilon)^2$,
which also cannot hold in the limit.
\end{proof}

Returning to the estimation of $k$ as a function of $S$,
consider the function $S \colon \mathbb{N} \to \mathbb{N}$,
defined as $S(k)=(2\sum_{i=1}^k p_i) - 2k + 2$.
Next, let $S(k)$ be extended to a real-valued function $f \colon \mathbb{R} \to \mathbb{R}$
that is continuous and increasing,
and coincides with $S$ on positive integers.
Then $f(x)=(1+o(1)) x^2 \ln x$,
and, by Lemma~\ref{sum_of_k_primes_inversion_lemma},
$f^{-1}(y) \sim \sqrt{\frac{2y}{\ln y}}$.
Since $f^{-1}(S(k))=k$ for all integer $k \geqslant 1$,
then $k$ is equivalent to $\sqrt{\frac{2S}{\ln S}}$.

This shows that $p_{k+1}$ is of the order
$k \ln k
	\sim
\sqrt{\frac{2S}{\ln S}} \ln\sqrt{S}
	=
\sqrt{2} \sqrt{\frac{S}{\ln S}} \frac{1}{2} \ln S
	=
\frac{1}{\sqrt{2}} \sqrt{S \ln S}$.
Since $n-S < 2p_{k+1}$, the values of $S$ and $n$ are equivalent,
and hence $k \sim \sqrt{\frac{2n}{\ln n}}$,
while the above estimation is equivalent to
$p_{k+1} \sim \frac{1}{\sqrt{2}} \sqrt{n \ln n}$.
And the loss of such a large term in the exponent of the lower bound $2^{\frac{S}{2}+k-1}$
would be too much,
because it would reduce the exponent down to
$\frac{n}{2} - \frac{1}{\sqrt{2}} \sqrt{n \ln n} + k - 1
	\sim
\frac{n}{2} - \frac{1}{\sqrt{2}} \sqrt{n \ln n} + \sqrt{\frac{2n}{\ln n}} - 1$
making it less than $\frac{n}{2}$.

An acceptable loss in the exponent is $o(k)$, that is, $o(\sqrt{\frac{n}{\ln n}})$.
It will be proved that 
if the last prime $p_k$ is replaced with the largest prime that fits in $n$,
then the loss in the exponent will be limited to $o(\sqrt{\frac{n}{\ln n}})$.

Recall that $k=k(n)$ is the greatest number satisfying
$S=(2\sum_{i=1}^k p_i) - 2k + 2 \leqslant n$.
Formally, the cycle lengths are chosen as $\pi_i=p_i$ for $1 \leqslant i \leqslant k-1$,
and $\pi_k=p \geqslant p_k$ is the maximum prime satisfying $(2\sum_{i=1}^{k-1} p_i) + 2p - 2k + 2 \leqslant n$.
In this case, the negative term in the exponent
may be as large as twice the distance between two consecutive primes
(namely, between $p$ and the next prime after $p$, where $p$ is between $p_k$ and $p_k+p_{k+1}$).
The best of the known upper bounds on this distance
is the theorem by Baker et al.~\cite{BakerHarmanPintz},
which reads that, for every $x$ large enough,
there is always at least one prime between $x-x^{0.525}$ and $x$.
Let $x=\frac{1}{2}\big(n-(2(\sum_{i=1}^{k-1} p_i) - 2k + 2)\big)$.
Then $x \geqslant p \geqslant p_k$,
and hence for all sufficiently large $n$ the theorem of Baker et al.\ is applicable to $x$.
On the other hand, $x < p_k + p_{k+1}$.
Then, if there exists a prime that exactly equals $x$,
this will provide a regular expression of size exactly $n$.
And if there is no such prime,
then there exists a prime between $x-x^{0.525}$ and $x$,
and taking this prime as $p$
will give a regular expression of size at least
$n-2x^{0.525} = n - O(\sqrt{n \ln n}^{0.525}) = n-o(\sqrt{\frac{n}{\ln n}})$.

Since the values of $S$ and $k$
in the lower bound $2^{\frac{S}{2}+k-2}$ obtained above
are estimated as 
$k \sim \sqrt{\frac{2n}{\ln n}}$
and $S=n-o(\sqrt{\frac{n}{\ln n}})$,
the lower bound on the number of states in a DFA
takes the following form.
\begin{equation*}
	2^{\frac{S}{2}+k-2}
		\geqslant
	2^{\frac{n}{2} - o(\sqrt{\frac{n}{\ln n}}) + (1+o(1))\sqrt{\frac{2n}{\ln n}}}
		\geqslant
	2^{\frac{n}{2} + (1+o(1))\sqrt{\frac{2n}{\ln n}}}
\end{equation*}
This completes the proof of Theorem~\ref{NFA_remembering_last_symbol_to_DFA_lower_bound_theorem}.

\section{Conclusion}

It has been proved that recognizing languages
defined by regular expressions of alphabetic width $n$
by a DFA
requires at least $2^{\frac{n}{2}+(\sqrt{2} + o(1))\sqrt{\frac{n}{\ln n}}}$ states
and at most $2^{\frac{n}{2}+(\frac{\log_2 e}{2\sqrt{2}}+o(1))\sqrt{n\ln n}}$ states.
Therefore, the first term in the exponent, $\frac{n}{2}$ is known precisely,
and the second term is known up to the power of the logarithm: $\sqrt{n}(\log n)^{\Theta(1)}$.
Determining the exact power of this logarithm is an interesting open problem.

\end{document}